\documentclass[aps,prl,superscriptaddress,floatfix,preprint]{revtex4-1}
\usepackage{times}
\usepackage{graphicx}
\usepackage{amsmath}
\usepackage{amssymb}
\usepackage{subfigure}
\usepackage{color}
\usepackage{epstopdf}
\usepackage{tikz}
\usepackage[colorlinks,citecolor=blue,linkcolor=red]{hyperref}

\pdfoutput=1

\begin{document}

\title{Induced Ferromagnetism at BiFeO$_3$/YBa$_2$Cu$_3$O$_7$ Interfaces}

\author{Jian-Xin Zhu}
\affiliation{Theoretical Division, Los Alamos National Laboratory,
Los Alamos, New Mexico 87545, USA}
\affiliation{Center for Integrated Nanotechnologies, Los Alamos National Laboratory,
Los Alamos, New Mexico 87545, USA}

\author{Xiao-Dong Wen}
\affiliation{Theoretical Division, Los Alamos National Laboratory,
Los Alamos, New Mexico 87545, USA}

\author{J. T. Haraldsen}
\affiliation{Theoretical Division, Los Alamos National Laboratory,
Los Alamos, New Mexico 87545, USA}
\affiliation{Center for Integrated Nanotechnologies, Los Alamos National Laboratory,
Los Alamos, New Mexico 87545, USA}
\affiliation{Department of Physics and Astronomy, James Madison University, Harrisonburg, Virginia 22807, USA}

\author{Mi He}
\affiliation{Division of Physics and Applied Physics, School of Physical and Mathematical Sciences,
Nanyang Technological University, Singapore 637371, Singapore}

\author{C. Panagopoulos}
\affiliation{Division of Physics and Applied Physics, School of Physical and Mathematical Sciences,
Nanyang Technological University, Singapore 637371, Singapore}

\author{Elbert E. M. Chia}
\affiliation{Division of Physics and Applied Physics, School of Physical and Mathematical Sciences,
Nanyang Technological University, Singapore 637371, Singapore}

\begin{abstract}
{\bf \noindent
Transition metal oxides (TMOs) exhibit many emergent phenomena ranging from high-temperature superconductivity and giant magnetoresistance to magnetism and ferroelectricity. 
In addition, when TMOs are interfaced with each other,  new functionalities can arise, which are absent in individual components. Here, we report results from first-principles calculations on the magnetism at the BiFeO$_3$/YBa$_2$Cu$_3$O$_7$ interfaces. By comparing the total energy for various magnetic spin configurations inside BiFeO$_3$, we are able to show that a metallic ferromagnetism is induced near the interface. We further develop an interface exchange-coupling model and place the extracted exchange coupling interaction strengths, from the first-principles calculations, into a resultant generic phase diagram.  Our conclusion of interfacial ferromagnetism is confirmed by the presence of a hysteresis loop in field-dependent magnetization data. The emergence of interfacial ferromagnetism should have implications to electronic and transport properties.
}
\end{abstract}
\maketitle

{\it Introduction.~}  An interface between two different transition metal oxides (TMOs) can generate novel emergent states that are typically absent in its constituent TMO bulk. For example, when two TMO insulators LaAlO$_3$ and SrTiO$_3$ are used to form a bilayer heterostructure, a metallic state emerges at the interface. This kind of metallicity produced out of insulator is usually accompanied by the collective electronic phenomena, such as high-temperature superconductivity and colossal magnetoresistance. Therefore, the interface platform can provide a unique opportunity for the design of interesting and controllable collective electronic properties, which cannot be realized with the individual constituents, and are more versatile in functionality as compared to their semiconductor counterparts. Since the pioneering discovery made by Ohtomo and Hwang~\cite{AOhtomo:2004}, the research has immediately sparked a flurry of experimental and theoretical attempts to uncover other novel interfacial states. Among them, the potential to change the magnetic properties in the appropriately synthesized magnetic TMO heterostructures is of particular interest. Examples include the observation of a magnetic coupling between La$_{0.67}$Sr$_{0.33}$MnO$_3$ layers in a La$_{0.67}$Sr$_{0.33}$MnO$_3$/YBa$_2$Cu$_3$O$_{7-\delta}$/La$_{0.67}$Sr$_{0.33}$MnO$_3$ superlattice~\cite{PPrzyslupski:2004} and a depression of the saturated magnetization in the La$_{0.67}$Ca$_{0.33}$MnO$_3$/YBa$_2$Cu$_3$O$_{7-\delta}$ superlattices~\cite{PPrieto:2001,JStahn:2005,AHoffmann:2005,JChakhalian:2006,JChakhalian:2007,WLuo:2008,ZLZhang:2009,JHoppler:2009,CVisani:2011,DKSatapathy:2012}, the  enhancement of magnetoelectric (ME) coupling  in some  piezoelectric/ferroelectric heterostructures such as 
CoFe$_2$O$_4$/BaTiO$_3$~\cite{HZheng04}, and also the induction of ferromagnetism 
in the antiferromagnet  BiFeO$_3$ (BFO) at the interface with ferromagnetic La$_{0.7}$Sr$_{0.3}$MnO$_{3}$~\cite{PYu10,SMWu10,SSingh:2013}. The BFO/ La$_{1-x}$Sr$_{x}$MnO$_{3}$ has recently been proposed as a candidate to engineer three-dimensional topological insulators~\cite{TDas:2013}. The control of interfacial ferromagnetism has also been demonstrated in multilayers where an insulating antiferromagnet is sandwiched between two ferromagnetic layers~\cite{JWSeo:2010}.   Technologically, this kind of research becomes especially relevant when one wants to control magnetization through the application of an electric field, which is mediated by ME coupling\cite{JHaraldsen:2013}. An intrinsic ME coupling is expected to occur most naturally in the so-called single-phase multiferroic materials, where both the time-reversal and space-inversion symmetries are absent.  The candidates for the multiferroic effect are BFO and $R$MnO$_3$ (where $R$ for rare earth elements). However, none of the existing single-phase multiferroic materials demonstrates significant and robust electric and magnetic polarizations at room temperature. In particular, materials like BFO and TbMnO$_3$ exhibit either commensurate or sinusoidal antiferromagnetism, this undesired property makes their potential technological applications limited. Therefore, the tuning of magnetism of these multiferroic materials when placed in contact with other TMOs is one of the most recent and fascinating research topics.

As already mentioned above, a variety of fascinating properties have been revealed in YBa$_2$Cu$_3$O$_{7-\delta}$-based junctions.  It is naturally anticipated that the BFO/YBa$_2$Cu$_3$O$_{7-\delta}$ heterostructures could give rise to novel properties~\cite{CLLu:2010}.
In this work, we focus on the interfacial magnetism in a BFO/YBa$_2$Cu$_3$O$_7$ (YBCO) superlattice. Within the first-principles density functional theory, we show that the ferromagnetism is induced near the interface in BFO. We further develop an interface exchange-coupling model to obtain a generic phase diagram.  By placing the magnetic exchange interaction strength extracted from the first-principles calculations, into the theoretical phase diagram, we are able to consolidate the observation of interfacial ferromagnetism in this composite material.

\begin{figure}[h!]
\centering\includegraphics[
width=1.0\linewidth,clip]{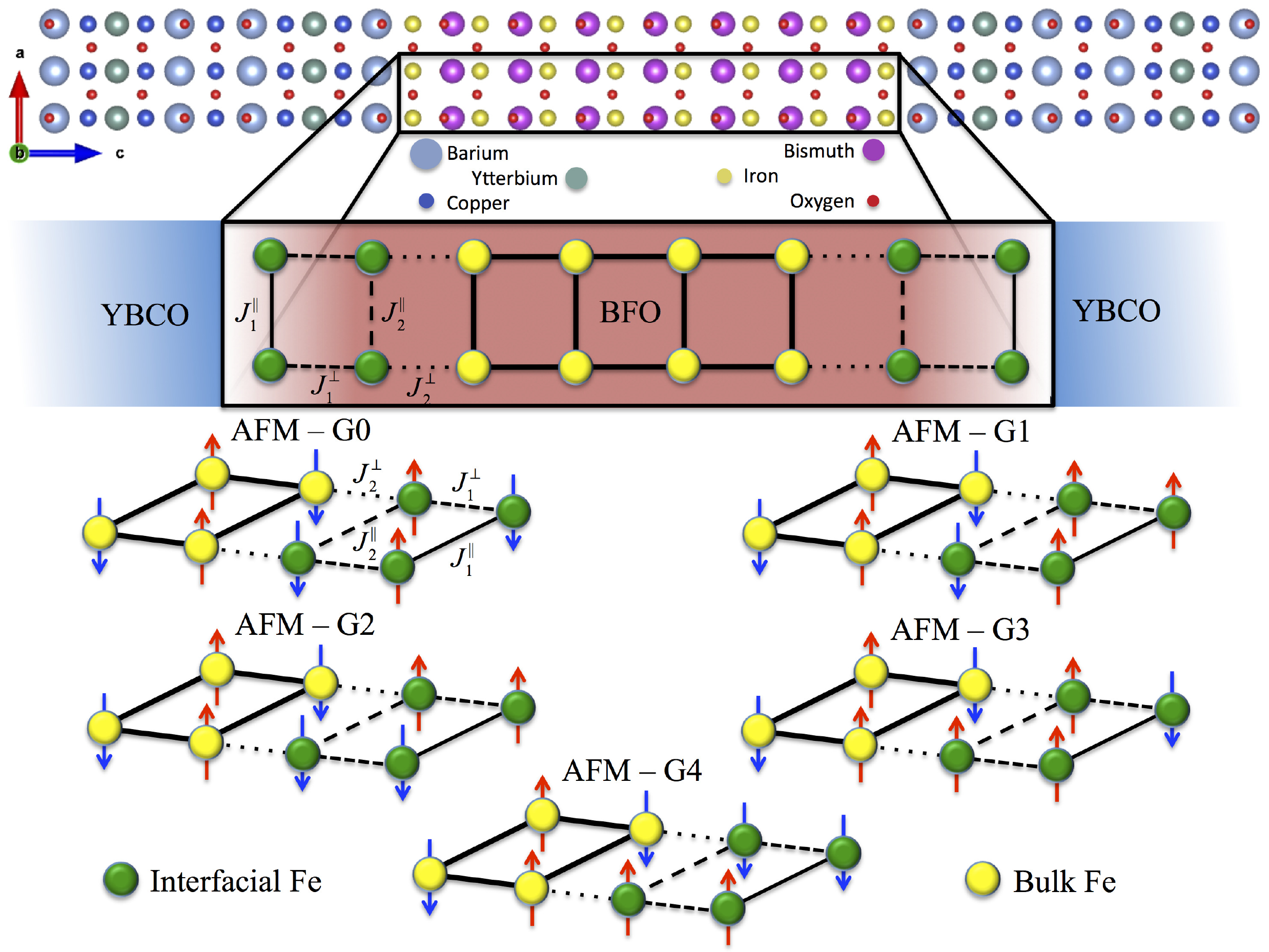}
\caption{(Color online) The crystallographic representation for the BFO/YBCO structure shown along the $ac$-plane (top panel). Below it is a schematic drawing of the Fe interactions within BFO/YBCO superlattice structure and five representative spin configurations on the Fe atoms of BFO. The spin configuration AFM-G0 is the same as that for the BFO bulk, for which a two-sublattice structure in the G-type AFM state is formed. For the spin configurations AFM-G1 through AFM-G4, the spin alignment deviates from that of the AFM-G0 in the first two layers of BFO near the interface. 
In the first-principles simulations, the following lattice constants of the supercell are used: $a=b=5.564\; \AA$    and $c=51.282\; \AA$.}
\label{fig:lda}
\end{figure}

{\em First-principles simulations.~}  The {\em ab initio} calculations are performed based on the density functional theory by using the plane-wave basis set and the projector-augmented-wave method~\cite{PEBlochl94} as implemented in the Vienna simulation package (VASP) code~\cite{GKresse96}. Calculations are carried out within the local spin-density approximation to the exchange-correlation functional plus on-site Hubbard repulsion (LSDA + $U$) on $d$-orbitals of Fe.  As in Ref.~\onlinecite{WLuo:2008}, no repulsion is introduced for the $d$-orbitals of Cu atoms when the YBCO is in the optimally doped regime. The band renormalization effect, due to the electronic correlation in the paramagnetic state, should be cancelled when one looks into the relative energy among various spin configurations in BFO. We choose a fixed value of $U_{eff} = 4$ eV on Fe 3$d$ orbitals throughout the work. A 500 eV energy cut-off was used to ensure the convergence of the total energy to 0.01 meV. For the BFO/YBCO superlattice under consideration, the supercell consists of 8 layers of BFO and 2 units of YBCO. In each unit of YBCO, there are 1 CuO chain and 2 CuO$_2$ planes along the stacking direction, except that the interfacial CuO chain in YBCO  is missing and replaced by FeO$_2$ plane (an assumption followed from the scanning transmission electron microscopy analysis on LCMO/YBCO superlattices~\cite{MVarela:2003}); while the lattice constant in each plane is chosen such that the in-plane sublattice structure for the G-type antiferromagnetic (AFM) state in BFO is accommodated. The Brillouin zone was sampled through a mesh of $4 \times 4 \times 1$ $k$-points. Due to the computational cost, all calculations are performed with the perfect superlattice without further atomic position relaxation of the structure. Since our focus in this work is on the magnetic structure rather than the ferroelectricity, we do not expect a significant change from the atomic position relaxation. This assumption seems to be supported by the electron microscopy observation on other TMO interfacial materials that the atomic displacements near the interface are less than 0.1 \AA~\cite{MVarela:2003}. 
Furthermore, since the BFO has a collinear G-type antiferromagnetism while the normal state of YBCO is non-magnetic, we consider only collinear spin polarization for various possible spin configurations on Fe atoms, as schematically depicted in Fig.~\ref{fig:lda}. The reason to consider various spin configurations lies in the fact that the density functional theoretical calculations cannot automatically find out the global ground state when different spin states are too close in energy. The self-consistency calculations are iterated until the energy difference between two consecutive iterations is less than $1\times 10^{-5}$ eV.

\begin{figure}[h!]
\centering
\includegraphics[
width=0.75\linewidth,clip]{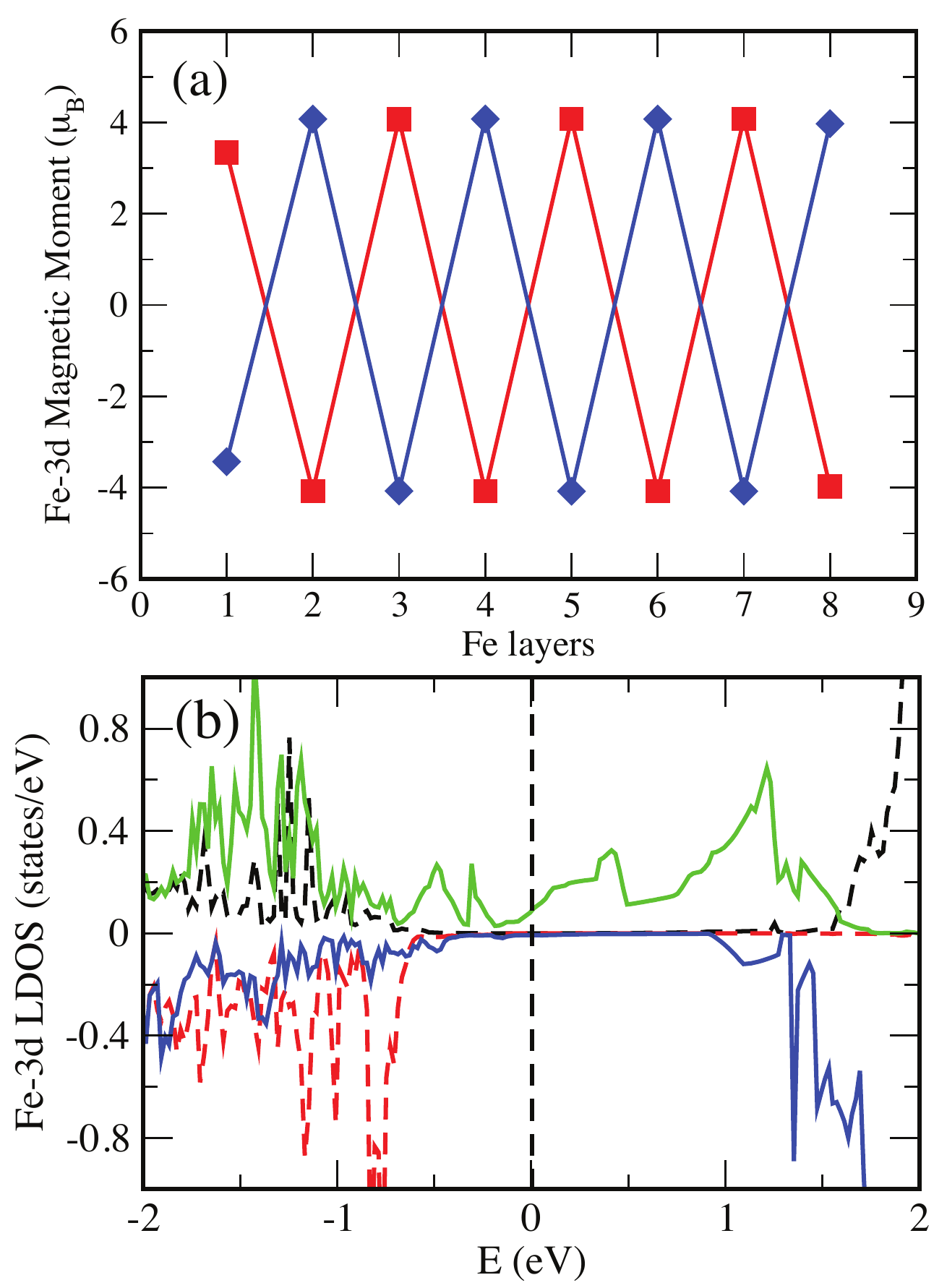} 
\caption{(Color online) Magnetic moment distribution contributed from Fe 3$d$ electrons in the BFO/YBCO superlattice obtained for the AFM-G0 spin configuration  (a) and the local density of states for Fe 3$d$ orbitals in the first layer of BFO near the interface (solid lines) and deep into the BFO segment (dashed lines) obtained from the AFM-G1 spin configuration (b). In (a), alternating moments are indicating the G-type AFM in the two sublattices in each FeO$_2$ plane.
}
\label{fig:mag_ldos}
\end{figure}

In Table~\ref{Table:Energy}, we show the relative energy for the five spin configurations corresponding to those described in Fig.~\ref{fig:lda}. As can be seen, the spin configuration AFM-G1, for which the spins on the Fe sites are aligned ferromagnetically in the first layer of BFO near the interface, has the lowest total energy. This suggests strongly the possibility of ferromagnetism emerging near the BFO/YBCO interface, although only limited number of spin configurations are considered within the first-principles calculations. To prove that this spin configuration is indeed the globally  stable state in energy, it is necessary to consider this spin configuration in the context of global phase diagram for the BFO/YBCO superlattice structure, which will be discussed immediately below. In addition, the self-consistent results from the first-principles method shows only a slight reduction of Fe-3$d$ magnetic moments at the interface, as demonstrated in Fig.~\ref{fig:mag_ldos}(a) for the AFM-G0 spin configuration  while the induced magnetic moment on Cu atoms (not shown here) in YBCO near the interface is only at the order of 0.01 $\mu_{B}$. The slight asymmetric interfacial magnetic moment is due to the uniaxial shift of O atoms a little away from the Bi planes. These theoretical results suggest the change in magnetic properties occurs with one or two layers of BFO away from the interface. This kind of short-ranged electron evolution has been revealed by cross-sectional scanning tunneling microscopy (STM) on cuprate/manganite interfaces~\cite{TYChien:2013}. Furthermore, we also show in Fig.~\ref{fig:mag_ldos}(b) the local density of states (LDOS) for Fe 3$d$ orbitals inside the BFO segment. Interestingly, although the Fe 3$d$ states are localized deep into the BFO segment, those Fe 3$d$ electrons of BFO near the interface are in the metallic state, which is readily accessible to an experimental test by the cross-sectional STM technique. By checking the electron charge from the first-principles simulations, we see that the total valence electronic charge on the Fe atoms nearest to the interface is larger than those that are deeper into BFO by an amount of about 0.2. It seems to suggest the charge transfer plays an important role in the emergence of ferromagnetism and metallic electronic state at the interface.

\begin{figure}[h!]
\centering\includegraphics[
width=0.8\linewidth,clip]{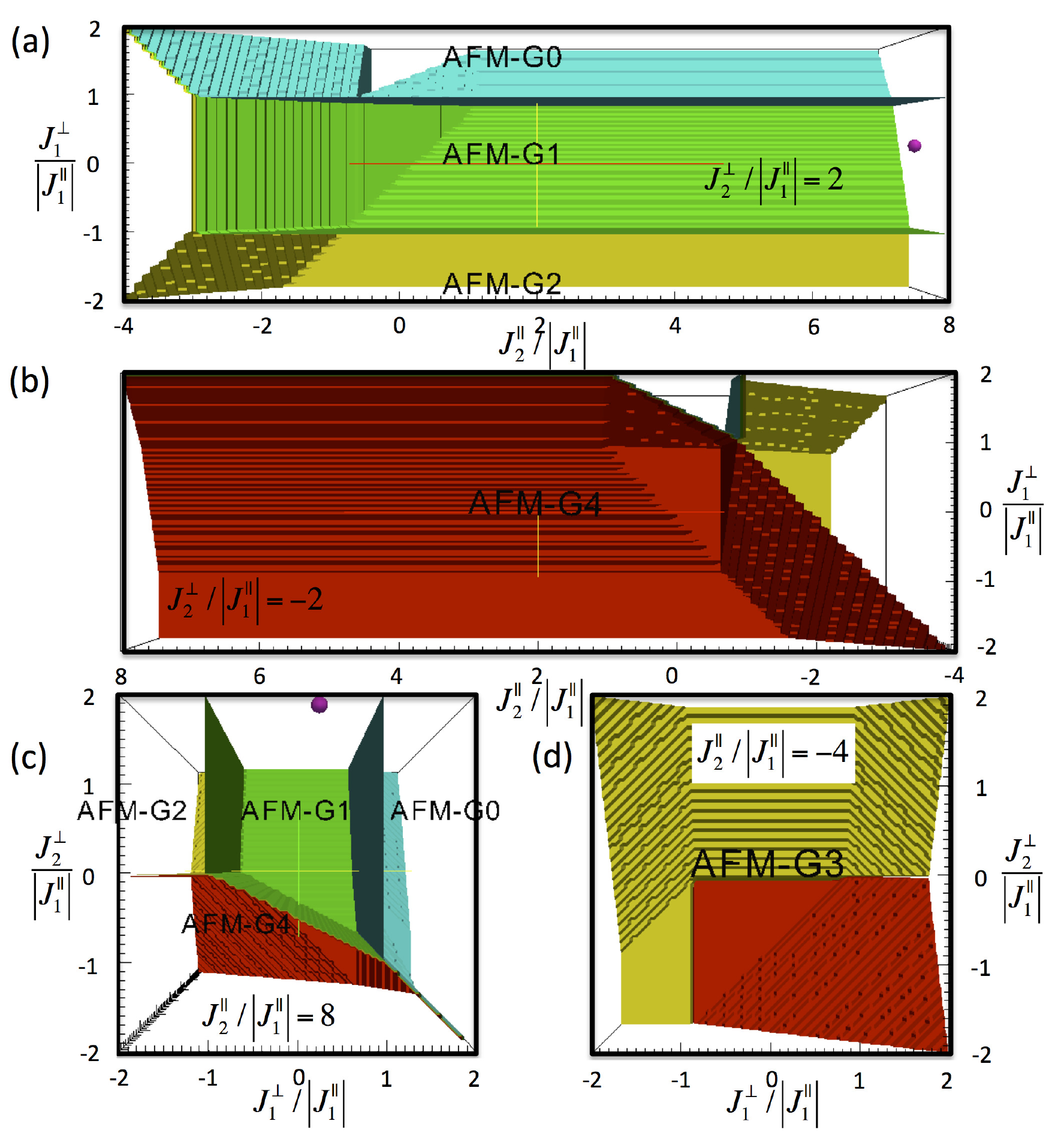}
\caption{(Color online) The 3D phase diagram for the normalized parameters based on the spin-exchange model for the five spin configurations of the BFO/YBCO superlattice structure assuming $J_{1}^{||}$/$|J_{1}^{||}|$ = -1. (a)  $J_{1}^{\perp}$/$|J_{1}^{||}|$ versus $J_{2}^{||}$/$|J_{1}^{||}|$ as viewed from the $J_{2}^{\perp}$/$|J_{1}^{||}|$ = 2 plane.  (b)  $J_{1}^{\perp}$/$|J_{1}^{||}|$ versus  $J_{2}^{||}$/$|J_{1}^{||}|$  as viewed from the $J_{2}^{\perp}$/$|J_{1}^{||}|$ = -2 plane.
 (c)  $J_{2}^{\perp}$/$|J_{1}^{||}|$ versus $J_{1}^{\perp}$/$|J_{1}^{||}|$ as viewed from the $J_{2}^{||}$/$|J_{1}^{||}|$ = 8 plane. 
 (d)  $J_{2}^{\perp}$/$|J_{1}^{||}|$ versus $J_{1}^{\perp}$/$|J_{1}^{||}|$ as viewed from the $J_{2}^{||}$/$|J_{1}^{||}|$ = -4 plane. 
The colors represent the borders between configuration regions (not the regions themselves). The blue border confines the AFM-G0 region from the AFM-G1, AFM-G3, and AFM-G4 regions. The green border confines the AFM-G1 region from all other regions. The yellow border confines the AFM-G2 region from the AFM-G1, AFM-G3, and AFM-G4 regions. The red border confines the AFM-G4 region from all other regions. The purple dot in panels (a) and (c) denotes the ground state parameters determined from the total energy calculations through the density functional theory. This shows that AFM-G1 is the ground state spin configuration. 
}
\label{fig:ex_cpl}
\end{figure}

{\em Effective exchange-coupling modeling.~}
The only slight reduction of the Fe 3$d$ magnetic moments near the interface of BFO/YBCO superlattice justifies an analysis of the interfacial magnetism within an effective spin-exchange model. Since the change in the magnetism occurs in only one or two layers of BFO away from the interface, the position dependent exchange couplings are restricted to these two layers. To investigate the various spin states for the BFO interface, we evaluate the spin configurations by evaluating the zeroth order spin exchange Hamiltonian within a classical limit~\cite{hara:09,JXZhu:2010}. This gives
\begin{equation}
\begin{array}{ll}
E &\displaystyle = \frac{1}{2}\sum_{i \neq j} J_{ij} \mathbf{{S}}_i \cdot \mathbf{{S}}_j \\
& \displaystyle = \frac{1}{2}\sum_{i \neq j} J_{ij} S^2 \cos(\theta_i-\theta_j)\;,
\label{genH}
\end{array}
\end{equation}
where $J_{ij}$ is the exchange parameter between spins $i$ and $j$, $S$ is the spin of the system, and $\theta_i-\theta_j$ is the difference between the spin orientations. Since the spins in these systems are collinear, $ \cos(\theta_i-\theta_j)$ is reduced to either 1 (0$^{\circ}$ or ferromagnetically aligned) or -1 (180$^{\circ}$ are antiferromagnetically aligned). Table \ref{Table:Energy} shows the calculated classical energy for each spin configuration as well as the interfacial spin configuration. By solving the above model Hamiltonian, we plot in Fig.~\ref{fig:ex_cpl} to show the three-dimensional (3D) phase diagram for the spin configurations  in the exchange parameter space as normalized by the parameter $|J^{||}_1|$. The phase diagram is represented in 3D to show the depth for the spin configuration regions. Although there are other possible spin configurations, this study focuses on the most likely candidates based on the total energy calculated by density functional theory. 
By fitting the classical energies for each phase with the density functional theory determined total energies, we can extract the exchange parameters $J_{1}^{||}$ = -0.02302 eV, $J_{2}^{||}$ = 0.17329 eV, $J_{1}^{\perp}$ = 0.0061225 eV, and $J_{2}^{\perp}$ = 0.04502 eV. 
By placing this set of parameter values normalized to $|J^{||}_1|$ into the phase diagram, we can establish the ground state phase, which is shown in Fig. \ref{fig:ex_cpl}(a) and (c) by the purple dot. This demonstrates that AFM-G1 is the ground state phase, and provides the first details that the interfacial spins in BFO are ferromagnetically aligned because of the electron interactions with YBCO. 

\begin{figure}[h!]
\centering\includegraphics[
width=0.50\linewidth,clip]{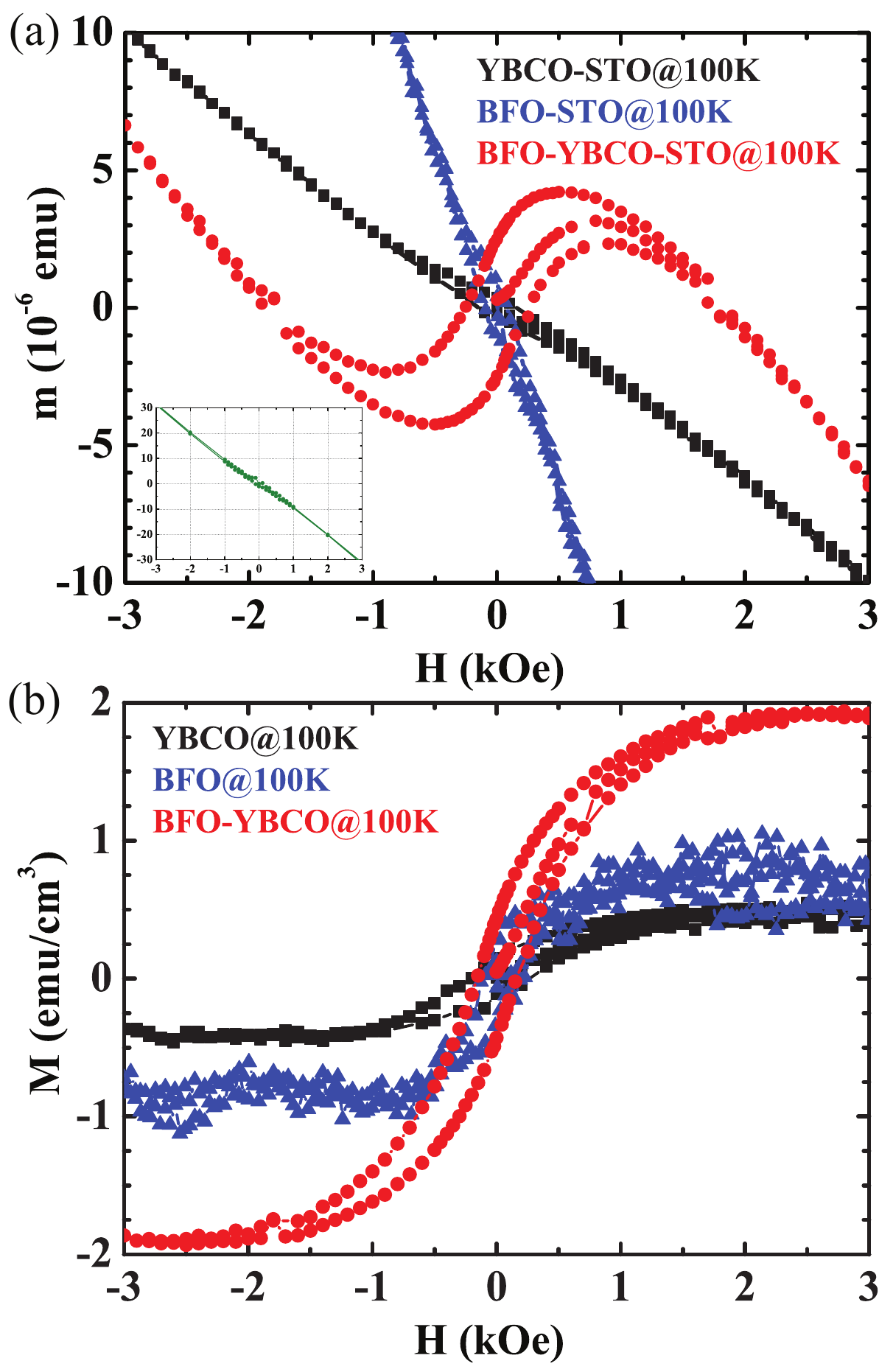}
\caption{(Color online) Field dependence of the magnetic moment (a) and magnetization (b) in the BFO/YBCO bilayer structure (red circles) as well as individual BFO (blue triangles) and YBCO (black squares) thin films deposited on a SrTiO$_3$ substrate.  The data are collected at 100 K, which is above the superconducting transition temperature ($\sim90$ K) of YBCO bulk. To obtain the magnetization, as shown in panel (b) for each individual sample, we subtract the magnetic contribution from the SrTiO$_3$ substrate (see the inset to panel (a) for an example) by assuming a linear background throughout the entire magnetic-field range. The small hysteresis loop observed in the YBCO sample is due to a weak ferromagnetism in the SrTiO$_3$ substrate possibly produced by oxygen vacancies~\cite{WXu:2013}.
}
\label{fig:exp_mag}
\end{figure}

{\em Magnetization measurement.~} We have also performed the SQUID measurements on the BFO/YBCO bilayer structure as well as separate BFO ($\sim$130 nm) and YBCO ($\sim$100 nm) thin films on a SrTiO$_3$ substrate. Figure~\ref{fig:exp_mag} shows the field dependence of the magnetic moment and magnetization on the three structures at 100 K, at which the YBCO is in the normal state while the BFO bulk is already in the AFM state ($T_N \sim 643$ K). A hysteresis characteristic of ferromagnetism in the BFO/YBCO bilayer, as shown in 
Fig.~\ref{fig:exp_mag}(a), is clearly seen. We note that the data of magnetic moment in Fig.~\ref{fig:exp_mag}(a) includes a tiny diamagnetic contribution from the SrTiO$_3$ itself, which is demonstrated representatively in the inset. Once we subtract this contribution within a linear background approximation, and obtain the magnetization only for YBCO, BFO, and BFO/YBCO thin films, the magnetic characteristics as shown in Fig.~\ref{fig:exp_mag}(b) become more indicative.  For our 130 nm thick BFO thin film, we do see a weak ferromagnetism (though the magnetization data is a little noisy). This result is not inconsistent with the early observation that the magnetic properties in BFO thin films are thickness dependent~\cite{JWang:2005}.  In contrast to the case of YBCO and BFO thin films, the magnetization for the BFO/YBCO bilayer structure exhibits a strong hysteresis loop, suggesting the ferromagnetic induction.  
 Similar hysteresis loop has also been observed at 150 K for the BFO/YBCO bilayer structure fabricated on [(LaAlO$_3$)$_{0.3}$(Sr$_2$TaAlO$_6$)$_{0.7}$] substrate~\cite{KWerner-Malento:2009}.  These experimental observations fully support the theoretical results obtained in the present work. 

{\em Conclusion.}   
In summary, we have explored the magnetic properties of BFO/YBCO superlattice structure within the density functional theoretical method.  Our total-energy calculations together with the effective spin-exchange modeling have enabled to show that although the BFO bulk is G-type AFM, there existence of FM in the BFO near the interface. Our theoretical prediction has then been further supported by the magnetization measurement in the BFO/YBCO bilayer structure. 

{\em Acknowledgments.} 
One of the authors (J.-X.Z.) thanks W. Luo for helpful discussions, and is also grateful to J. Qi, R. Prasankumar, and Q. Jia for collaboration on related topics. This work was supported by U.S. DOE at LANL under Contract No. DEAC52-06NA25396 and LANL LDRD-DR Program (J.-X.Z., X.D.W., \& J.T.H.), Singapore NRF-CRP-2008-04 (M.H., C.P., \& E.E.M.C.) and MOE AcRF Tier 1 RG 13/12 (E.E.M.C.). This work was also, in part, supported by the Center for Integrated Nanotechnologies, a U.S. DOE Office of Basic Energy Sciences user facility.

\vspace{0.5cm}
{\noindent\large\bf Author contributions}\\
J.-X.Z. conceived and designed the study, and carried out with X.D.W. the numerical calculations. 
J.T.H. carried out the theoretical modeling. M.H., C.P., and E.E.M.C. performed the experiment. All authors contributed to the scientific discussions. J.X.Z. and J.T.H. wrote the paper with inputs from the rest of co-authors.

\vspace{0.5cm}
{\noindent\large\bf Additional Information}\\
{\noindent\bf Competing financial interests:} 
The authors declare no competing financial interests.

{\noindent\bf Reprints and permission}
is available
online at http://npg.nature.com/reprintsandpermissions/ \\
Correspondence and requests for
materials should be addressed to J.-X.Z. (email: jxzhu@lanl.gov).


\newpage
\noindent {\bf FIGURE LEGENDS}

\begin{table}[h!]
\caption{Classical energies and the relative energies $\Delta E$ (eV/supercell) for five spin configurations in the BFO.}
\begin{ruledtabular}
\begin{tabular}{lccc}
Spin Config. & Interface Spins & $(E_i-E_0)/S^2$ & $\Delta E$ (eV)\\
\hline \\
AFM-G0 &  AFM & 0 & 0\\
AFM-G1 &  FM & 4$J^{||}_{1}$+4$J^{\perp}_{1}$ & -0.06759\\
AFM-G2 &  AFM & 8$J^{\perp}_{1}$ &  0.04898\\
AFM-G3 &  AFM & 4$J^{||}_{2}$+4$J^{\perp}_{1}$+4$J^{\perp}_{2}$ & 0.89778 \\
AFM-G4 &  AFM & 8$J^{\perp}_{1}$+8$J^{\perp}_{2}$ & 0.40919\\
\label{Table:Energy}
\end{tabular}
\end{ruledtabular}
\end{table}

\end{document}